# First principles investigations of the electronic and magnetic properties of $RVO_3$ (R = Er, Ho, Y, Lu) vanadates

S. Ait Jmal[1*], M. Balli[1*], H. Bouhani[2], O. Mounkachi[3,4]

[1] AMEEC Team, LERMA, International University of Rabat, Parc Technopolis, Rocade de Rabat- Salé, 11100, Morocco.

[2] Complex Systems and Interactions, Ecole Centrale Casablanca, Morocco

[3] Laboratory of Condensed Matter and Interdisciplinary Sciences, Department of Physics, Faculty of Sciences, Mohammed V University in Rabat, PO Box 1014, Rabat, Morocco.

[4] Modeling, Simulation & Data Analysis Program, Mohammed VI Polytechnic University, Ben Guerir 43150, Morocco.

*Contact: mohamed.balli@uir.ac.ma (MB), sohail.aitjmal@uir.ac.ma (S.A)

## Abstract

In this paper, we investigate the influence of the rare earth element (R) on the physical properties of $RVO_3$ (R = Er, Ho, Y, Lu). For this purpose, a theoretical work is reported in this study using ab initio method with the intent of enhancing our knowledge of the electronic, and magnetic properties of $RVO_3$ (R = Er, Ho, Y, Lu) compounds. The electronic properties of these materials show the presence of Jahn-Teller effect, with a band gap of approximately 0.58eV, 1.57eV, 1.12eV, 1.28eV, for $ErVO_3$, $HoVO_3$, $YVO_3$, $LuVO_3$, respectively. The magnetic contribution of Ho, Er, and V atoms in $HoVO_3$, and $ErVO_3$ compounds was disentangled using X-ray magnetic circular dichroism (XMCD). The magnetic anisotropies of $RVO_3$ with (R = Er, Ho, Y, Lu) are studied and analyzed using spin orbit coupling. We noticed an easy axis of magnetization along the c-direction with strong magneto-crystalline anisotropy energies for the non-magnetic rare earth elements (R = Lu, Y)$VO_3$. For the magnetic rare earth elements, the easy and hard magnetization direction were found to be along b-axis and c-axis, and a- axis and c-axis for $HoVO_3$, and $ErVO_3$, respectively. This means that instead of the standard magnetization demagnetization process, large thermal effects can be obtained by spinning their single crystals in constant magnetic fields. The exchange coupling between (Ho-V) is also found to be greater than the other exchange interactions in $HoVO_3$. Accordingly, a large conventional magnetocaloric effect could be created in $HoVO_3$, since the Ho Ising moments experience a metamagnetic transition. However, the magnetocaloric effect (MCE) is expected to be lower in ErVO$_3$ due to the weak (Er-V) exchange coupling.

## 1. INTRODUCTION

In our daily life, the temperature magnetic cooling systems, which are based on the intrinsic magnetocaloric effect (MCE) of magnetic materials, is a trending technology that attracts growing attention thanks to its eco-friendly nature and its massive potential at being energy efficient, compared to the traditional cooling system [1]. Generalizing the use of this alternative technology would lead to completely dispose of the harmful synthetic refrigerants [2], and thus achieving one of the major aims of many treaties that were universally adopted by the international community aiming to minimize toxic gas usage [3]. According to literature, the use of MCE in cooling systems at near and low temperatures is quite promising [4-7]. One of the main problems in developing a magnetic refrigerator is the search for a suitable working material with a high MCE which is characterized by various physical quantities such as a Relative Cooling Power (RCP), isothermal magnetic entropy change ($-\Delta S_{mag}$) and adiabatic temperature change ($\Delta T_{ad}$) [8].

Materials based on complex oxides of elements with empty d- and f-electron shells have attracted the attention of both researchers and industrials because of their unique combination of electric, magnetic, optical, ferroelectric, and other relevant properties [9-10]. Their specificity is largely determined by the degree to which vacant electron shells are filled, the lengths of the corresponding orbitals, and the nature of electron involvement in interatomic chemical bonds. The change in those factors causes a change in the conductivity in this type of oxides. As a result, members of this family of compounds can be classified not only as insulators and semiconductors, but also as conducting materials [11-13].

In this context, the RMO$_3$ series of transition metal oxide compounds (R=rare earth, M=transition metal.) have already attracted considerable interest from the materials science community due to their fascinating physical properties such as high temperature superconductivity and large magnetocaloric effect. Particularly, the magnetocaloric features of RMnO$_3$ manganites have sparked worldwide interest. [14-16]. A. A. Wagh et al [17], have reported a large magnetocaloric effect at low temperature in GdMnO$_3$ of about $-\Delta S_{mag} \sim$ 31.8J(kg/K) at T $\sim$ 7K and $\Delta T_{ad} \sim$ 10K at 19.5K for (0-80kOe) field variation along the c axis. Other manganite oxides with anisotropic MCE have also been reported, including HoMnO$_3$

[18], DyMnO$_3$ [19], YbMnO$_3$ [20], TbMnO$_3$[21], and ErMnO$_3$ [22]. Unlike manganite oxides RMnO$_3$, only few studies have been reported on the magnetocaloric potential of RVO$_3$ vanadate oxides which are emerging materials for magnetocaloric cooling applications at low temperatures [23-24]. In this context, HoVO$_3$, ErVO$_3$, YVO$_3$, and LuVO$_3$ systems have attracted the attention of researcher. Particularly, because of their multiple fascinating physical properties such as orbital ordering, octahedral tilting, and Jahn-Teller effect [25].

The HoVO$_3$ undergoes a three phase transitions, which have been reported in the literature [25]. Two structural transitions from monoclinic to orthorhombic at T ~ 40 K, and from orthorhombic to monoclinic at T ~ 188K accompanied by G-type orbital ordering (OO), which depends mainly on the degree of octahedral tilting caused by the deviation in the size of the rare earth element [26]. HoVO$_3$ compound exhibits a Néel transition at T = 110 K due to the presence of antiferromagnetic C-type order of the vanadium sublattice.

M. Balli et al [28], showed that the HoVO$_3$ compound exhibits a large negative magnetocaloric effect of 7.7 (J/kg. K) for a field change of 1.4 T. In addition, the HoVO$_3$ compound shows a giant conventional MCE of 17.2 J/kg K for 7 T without thermal and magnetic hysteresis close to T$_{Ho}$. In addition, Bouhani et al. [27] showed that the magnetocaloric effect of PrVO$_3$ thin films reaches unusual maximum values of 56.8 J/kg K with the magnetic field change of 6 T applied in the sample plane over the cryogenic temperature range close to 3 K. On the other hand, S. Kumari et al. [29] have used GGA+U approximation to perform an ab-initio self-consistent computation of the electronic structure of LaVO$_3$ and YVO$_3$, revealing semiconducting behavior in the electronic band structure and density of states for both compounds with band gaps of 1.1 eV and 1.2 eV, respectively. However, to our knowledge similar investigations of RVO$_3$ with magnetic rare earth elements have not been reported up to know which is due more probably to their complicated structure and the lack of significant experiment findings such as band gap values.

In this paper, the Density Functional Theory (DFT) is used to investigate the structural, electronic, and magnetic properties of HoVO$_3$. A similar analysis is carried out to get insight on the difference between this compound and ErVO$_3$, as they have nearly identical structural behavior and magnetic moment, and therefore, explain the driven mechanisms behind their magnetocaloric properties. For comparison, we also looked at the structural, electronic, and magnetic properties of RVO$_3$ (R= Y, Lu) with a non-magnetic rare earth element.

## 2. COMPUTATIONAL DETAILS

The full-potential linearized augmented plane wave (FP-LAPW) method within the density functional theory (DFT) [30-33], as implemented in package WIEN2K [34], is utilized in our calculations. First, the popular generalized gradient approximation (GGA-PBE) [35] is used to optimize the crystal structure and investigate electronic structures and magnetism in $RVO_3$. Then GGA+U approximation with $U_{eff} = U - J = 2.85$ eV, and $U_{eff} = 4.36$ eV is used for V, and R atoms respectively [36,29]. Also, the spin-orbit coupling (SOC) is taken into consideration to calculate the magneto-crystalline anisotropy. To determine the later, DIPAN program implemented in WIEN2K code is used. The cut-off energy is set to -6.0 Ry to separate core states from valence states. The k-mesh size in the first Brillouin zone is 11x11x11. We set $R_{mt} \times K_{max}= 7$ and use magnitude of largest vector $G_{max} = 12$ in charge density Fourier expansion. The radii of Y, Lu, Er, Ho, V, and O atomic spheres are set to 2.26, 2.42, 2.4, 2.3, 1.85, 1.68 Bohr, respectively. The self-consistent calculations are converged only when the absolute charge-density difference par formula unit between the successive loops is less than 0.0001|e|, where e is the electron charge.

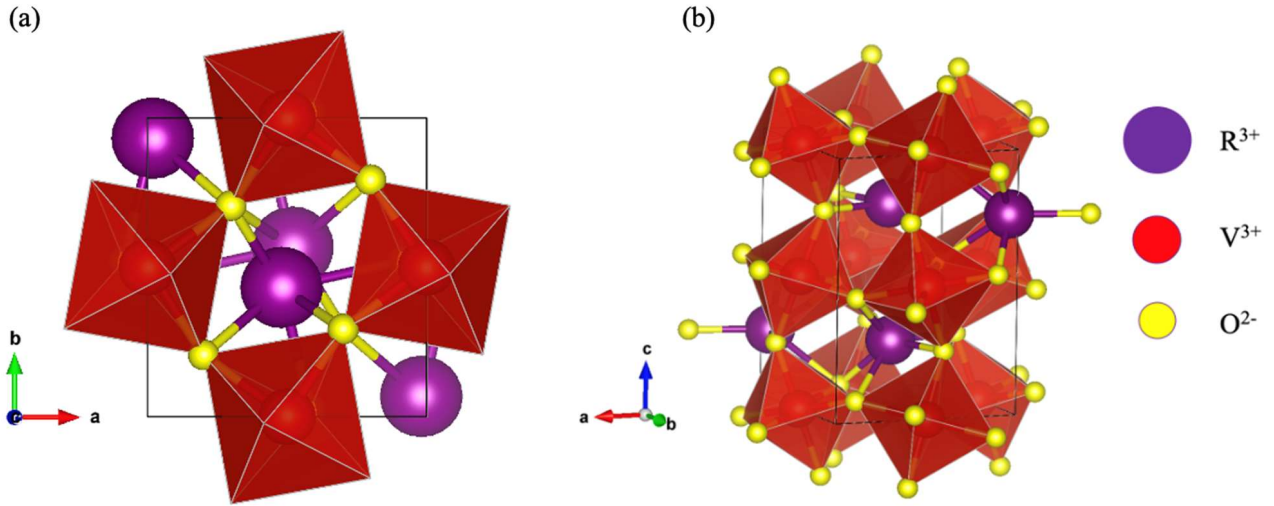

Figure 1: Crystal structure of $RVO_3$ orthovanadates. (a) top view, (b) side view. V atoms (red) surround the octahedron created by oxygen atoms (yellow spheres), whereas R atoms are represented by purple spheres. (Constructed with Vesta program [37]).

Theoretically, any substance having at least one nonmagnetic moment and a non-zero spin-orbit coupling may have its XMCD signal measured. The "sum rules" allow the study of XMCD spectra to return to the moment of spin and orbit of the examined $ErVO_3$, and $HoVO_3$ elements. The aim of the XMCD spectrum is to identify the orbital and spin moment, knowing that there is a link between the integral of XMCD spectrum and the mean value of the orbital angular momentum projection on the magnetization axis. Thole Carra and Van Der Laan [38] introduced this connection; subsequently, the same authors provided a second sum rule for spin moment [39].

$$\frac{\int_{j+} d\omega(\mu^+ - \mu^-)}{\int_{j\pm} d\omega(\mu^+ + \mu^- + \mu^0)} = \frac{c(c+1) - l(l+1) - 2}{2l(l+1)(4l+2-n)} \langle L_Z \rangle$$

$$\frac{\int_{j+} d\omega(\mu^+ - \mu^-) - c(c+1)\int_{j-} d\omega(\mu^+ - \mu^-)}{\int_{j\pm} d\omega(\mu^+ + \mu^- + \mu^0)} = \frac{l(l+1) - c(c+1) - 2}{3c(l+1)} \langle S_Z \rangle$$

Where:
- l = orbital quantum number of the valence state.
- c = orbital quantum number of the core state.
- $(\mu^-)(\mu^+)$ = absorption spectrum for right (left) circularly polarized light.
- $\langle T_Z \rangle$ = value of the magnetic dipole operator.

The sum rules enable us to extract individually the orbital magnetic moment $\langle L_Z \rangle$ and the spin magnetic moment $\langle S_Z \rangle$ [40] from the integrated XMCD. The value of the spin and orbital moments in ErVO3, and HoVO3 compounds are estimated using this signal. For the computation of orbital and spin moments, the XMCD spin moment sum rule is used to the $M_{4,5}$ edges of the rare earth element. For $M_{4,5}$ edges the transition 3d (c = 2) into 4f final states (l=3) with the number of holes $n = 4l + 2 - n_{4f}$:

$$m_L = (\mu_B/\hbar)\langle L_Z \rangle = -2((\int_{M4+M5} d\omega(\mu^+ - \mu^-)) / \int_{M4+M5} d\omega(\mu^+ + \mu^-)) * (14 - n_{4f})$$

$$m_s = (\mu_B/\hbar)\langle S_Z \rangle$$
$$= ((7\int_{M5} d\omega(\mu^+ - \mu^-)) - 6\int_{M5} d\omega(\mu^+ - \mu^-)) / (\int_{M4+M5} d\omega(\mu^+ + \mu^-)) * (14 - n_{4f})(1 + 10(\langle T_Z \rangle/\langle S_Z \rangle))^{-1}$$

where $L_Z$ and $S_Z$ are the orbital and spin moments of R atoms, respectively.

## 3. RESULTS AND DISCUSSION

### 3.1. Crystal structure

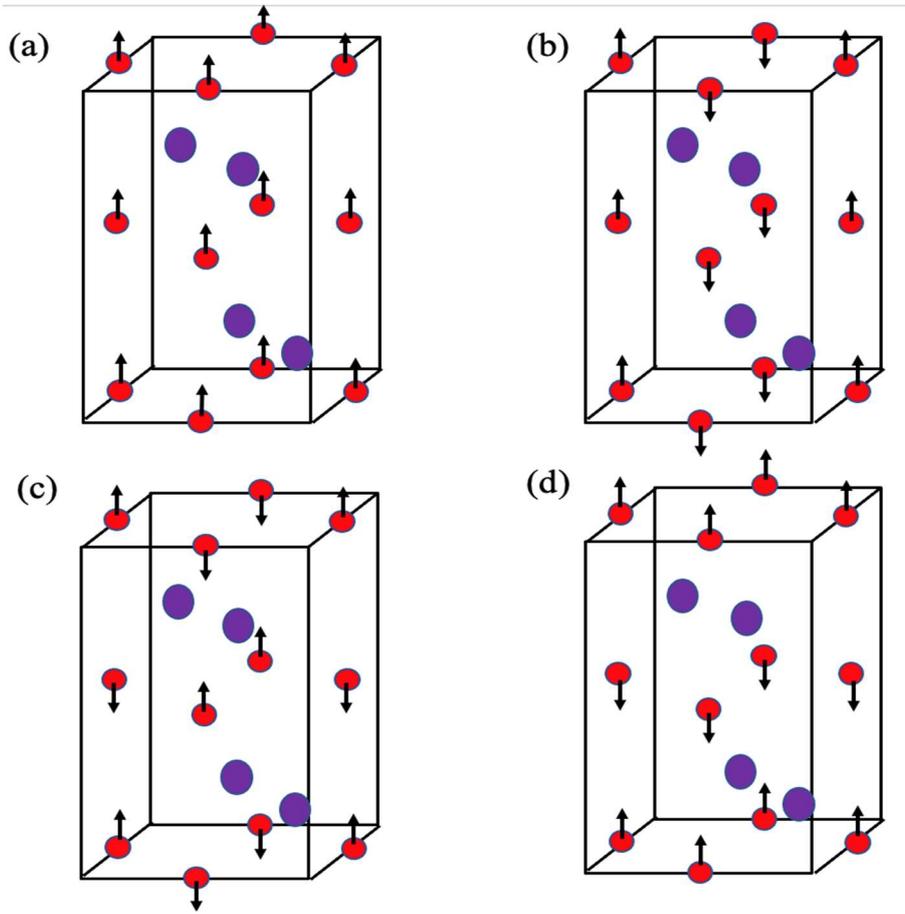

Figure 2: The magnetic structures are schematically shown: (a) ferromagnetic, (b) antiferromagnetic C-type, (c) antiferromagnetic G-type, and (d) antiferromagnetic A-type. The purple and red spheres represent R and V atoms, respectively. The Arrows indicate magnetic moment orientations of V atoms.

As shown in Fig.1, the orthorhombic $RVO_3$ investigated here has space group Pbnm (N° 62) [25] at low temperatures. First, the lattice parameters and ionic positions are optimized within GGA, and GGA+U approximations. The obtained results are summarized in Table 1. Available experimental results [41] are also presented for comparison. The GGA+U optimized atomic positions are summarized in Table 2.

The calculated values are well consistent with experimental lattices constants reported in [41].

Table 1: Optimized lattice parameters with GGA, GGA+U and experimental data for HoVO$_3$, ErVO$_3$, YVO$_3$, and LuVO$_3$.

| Compounds | | a | b | c |
|---|---|---|---|---|
| **HoVO$_3$** | GGA-PBE | 5.1943 | 5.2917 | 7.4084 |
| | GGA+U | 5.2190 | 5.4898 | 7.4910 |
| | Exp. [41] | 5.2853 | 5.5953 | 7.5891 |
| **ErVO$_3$** | PBE | 5.2644 | 5.5856 | 7.5716 |
| | GGA+U | 5.2712 | 5.6011 | 7.5980 |
| | Exp. [41] | 5.3696 | 5.6973 | 7.7230 |
| **YVO$_3$** | PBE | 5.3413 | 5.6298 | 7.7167 |
| | GGA+U | 5.3409 | 5.6320 | 7.6956 |
| | Exp. [41] | 5.2824 | 5.5927 | 7.5831 |
| **LuVO$_3$** | PBE | 5.3192 | 5.6271 | 7.7505 |
| | GGA+U | 5.3044 | 5.6120 | 7.7075 |
| | Exp. [41] | 5.2156 | 5.5611 | 7.58347 |

Table 2: Optimized atomic positions (x, y, z) with GGA+U for RVO$_3$ with (R = Ho, Er, Y, Lu) compared with experimental values.

|            | **Ho**      | **Er**      | **Y**       | **Lu**      |
|------------|-------------|-------------|-------------|-------------|
| **R**      |             |             |             |             |
| x (GGA)    | 0.980319    | 0.98279684  | 0.980523    | 0.98431936  |
| (Exp. [41])| (0.9811)    | (0.9798)    | (0.9808)    | (0.9788)    |
| x (GGA+U)  | 0.98013499  | 0.97261854  | 0.97773721  | 0.97679128  |
| y          | 0.064967    | 0.0647568   | 0.064964    | 0.0658614   |
| (Exp. [41])| (0.0682)    | (0.0694)    | (0.0682)    | (0.0708)    |
| y (GGA+U)  | 0.06949151  | 0.07130780  | 0.06896371  | 0.06913509  |
| z          | 0.25        | 0.25        | 0.25        | 0.25        |
| (Exp. [41])| (0.25)      | (0.25)      | (0.25)      | (0.25)      |
| z (GGA+U)  | 0.25        | 0.25        | 0.25        | 0.25        |
| **V**      |             |             |             |             |
| x          | 0.5         | 0.5         | 0.5         | 0.5         |
| (Exp. [41])| (0.5)       | (0.5)       | (0.5)       | (0.5)       |
| x (GGA+U)  | 0.5         | 0.5         | 0.5         | 0.5         |
| y          | 0           | 0           | 0           | 0           |
| (Exp. [41])| (0)         | (0)         | (0)         | (0)         |
| y (GGA+U)  | 0           | 0           | 0           | 0           |
| z          | 0           | 0           | 0           | 0           |
| (Exp. [41])| (0)         | (0)         | (0)         | (0)         |
| z (GGA+U)  | 0           | 0           | 0           | 0           |
| **O1**     |             |             |             |             |
| x          | 0.08170492  | 0.08076827  | 0.08170492  | 0.0815922   |
| (Exp. [41])| (0.1108)    | (0.1124)    | (0.1109)    | (0.1198)    |
| x (GGA+U)  | 0.11732973  | 0.12106600  | 0.11621836  | 0.11345226  |
| y          | 0.4782525   | 0.47926183  | 0.4781992   | 0.4772492   |
| (Exp. [41])| (0.4604)    | (0.4596)    | (0.4605)    | (0.4544)    |
| y (GGA+U)  | 0.46134336  | 0.45554690  | 0.45455470  | 0.46349289  |
| z          | 0.25        | 0.25        | 0.25        | 0.25        |
| (Exp. [41])| (0.25)      | (0.25)      | (0.25)      | (0.25)      |
| z (GGA+U)  | 0.25        | 0.25        | 0.25        | 0.25        |
| **O2**     |             |             |             |             |
| x          | 0.70260079  | 0.70547256  | 0.70341969  | 0.697129218 |
| (Exp. [41])| (0.6909)    | (0.6900)    | (0.6912)    | (0.6884)    |
| x (GGA+U)  | 0.68714866  | 0.69108397  | 0.69438844  | 0.68923910  |
| y          | 0.29111621  | 0.29040367  | 0.2891022   | 0.29359127  |
| (Exp. [41])| (0.3045)    | (0.3043)    | (0.3035)    | (0.3070)    |
| y (GGA+U)  | 0.30299822  | 0.30241134  | 0.30963101  | 0.30491752  |
| z          | 0.04319569  | 0.04326159  | 0.04802953  | 0.0481182   |
| (Exp. [41])| (0.0553)    | (0.0566)    | (0.05618)   | (0.06034)   |
| z (GGA+U)  | 0.05620842  | 0.06254009  | 0.05946116  | 0.05929415  |

The magnetic phase stability of RVO3 compounds with (R = Er, Ho, Y, Lu) are also examined by using antiferromagnetic A, C, and G-types in various configurations (see Fig. 2). (a, b, and c), and the ferromagnetic one (spins have the same direction). According to the obtained total energies results by GGA+U approximation, it was found that the stable magnetic configuration is C-type antiferromagnetic of vanadium ions with an antiferromagnetic configuration of R ions along the c-axis for ErVO3, HoVO3, YVO3, and A-type antiferromagnetic of vanadium ions with an antiferromagnetic configuration of Lu ions along the c-axis for LuVO3 compound. The other energies are higher. These findings are in accordance with experimental data. [41,42].

### 3.2. Electronic properties

To determine the nature of the electronic properties of RVO3 (R = Ho, Er, Y) C- type AFM and A-type AFM of LuVO3, the electronic band structure and density of state in the vicinity of Fermi energy ($E_F$) are calculated by using the ab-initio density functional theory (DFT) within GGA and GGA+U approximations. The effective interaction $U_{eff} = U - J$ has been set to 4.36eV for R atoms [36,29], and 2.85 eV for V atoms. Fig. 3(a-d) shows the total and partial density of state of HoVO3, ErVO3, YVO3, and LuVO3 using GGA approximation. Because the AF order is known to occur in this molecule, the computation was done using Wien2k's polarized spins capability. The GGA+U computations validate this AF order. The total magnetic moments of these materials are almost negligible ($0.01\mu_B$). In the absence of local Coulomb repulsion on rare earth and vanadium atoms, the band structure depicted in Figure 3(a-d) for all materials investigated here shows a metallic behavior, as is well known. Because the bands are not completely filled (the band created by d-orbital of vanadium electrons is half filled), the GGA approximation forecasts a metallic state because of the difficulty to treat strongly coupled systems due to underestimating electron-electron interactions. According to band theory [43], RVO3 compounds should be metals. However, they were experimentally found to be insulating due to significant onsite Coulomb repulsion [44]. For this purpose, GGA+U approximation was employed with U = 2.85 eV for vanadium 3d orbitals. The total DOS and band structure have significantly affected due to the inclusion of Hubbard U term, can be clearly seen in Figures 3(a-d) and 4(a-d). In Fig.(3a), electron-electron correlations resulting from the Coulomb repulsion allow us to identify the exact location of V (3d-$t_{2g}$) lower-Hubbard band (LHB) and upper-Hubbard band (UHB). The band gap which corresponds to the energy difference between these two LHB and UHB is found to be 1.58eV for HoVO3, being similar to the observed behavior in other oxides such as YTiO3, CaMnO3, LaFeO3 [45]. According to Fig.4 (b-d), gaps of 0.58 eV, 1.1 eV, and 1.28 eV are obtained for ErVO$_3$, YVO$_3$, and LuVO$_3$, respectively. This is confirmed by the reported band structures in Fig. 6(a-d). The forbidden gap, which divides the electron states in the valence and conduction bands as shown in Fig. 5(a-d), has a V 3d$_z^2$ – V 3d$_{xy}$ character in holmium vanadate and a V 3d$_{xz}$ – V 3d$_{yz}$ character in the ErVO$_3$ compound. A similar phenomenon was observed in other RVO$_3$ materials, such as LaVO$_3$ [46], and SrVO$_3$ [47] compounds.

It is worth noting that the $d^2$ configuration of transition metals is known to split into two sub levels in the octahedral crystal field. Fig.5 (a-d) shows the partial density of vanadium d orbitals, with the 3d levels separated into low energy t$_{2g}$ orbitals and high-energy $e_g$ orbitals. The separation of $t_{2g}$ level into two levels is attributed to the Jahn Teller effect [48]. This, along with other structural distortions, contributes to the promotion of the antiferromagnetic ground state.

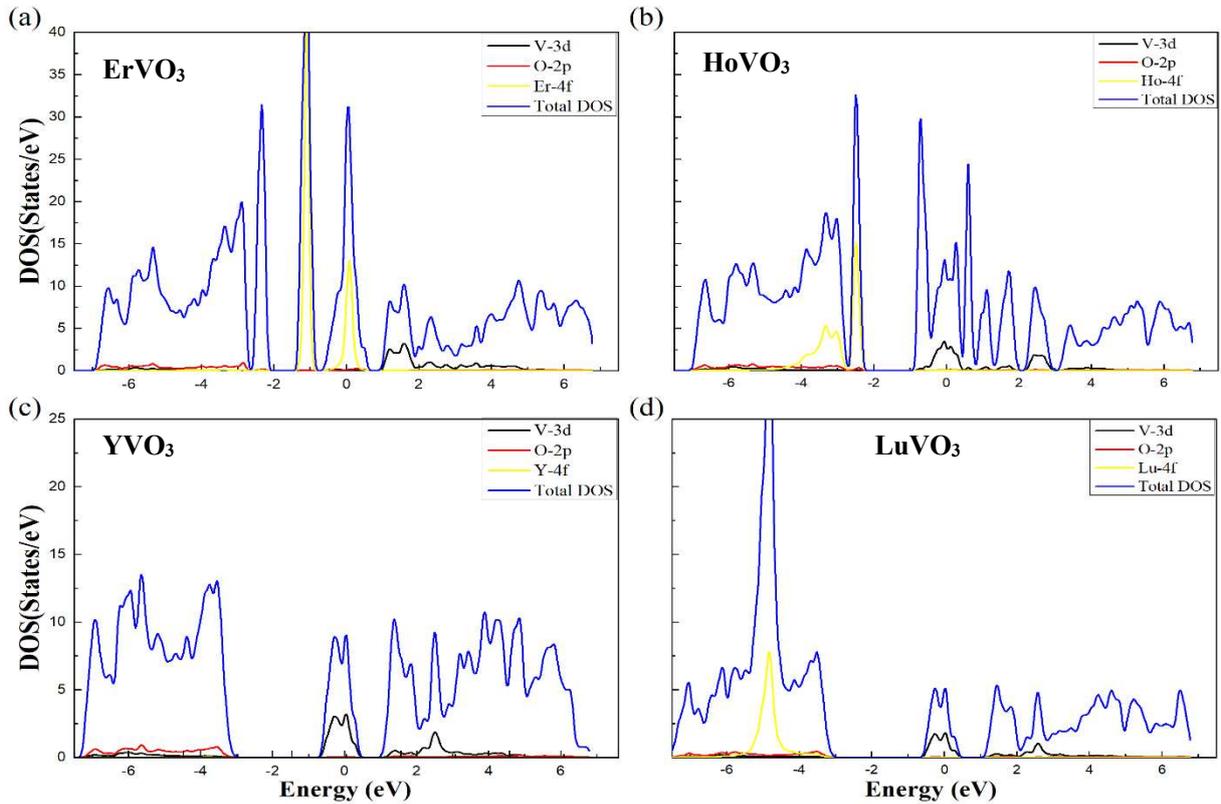

Figure 3: Total and partial density of states obtained within GGA of. (a) ErVO$_3$ (b), HoVO$_3$ (c), YVO$_3$ (d) LuVO$_3$.

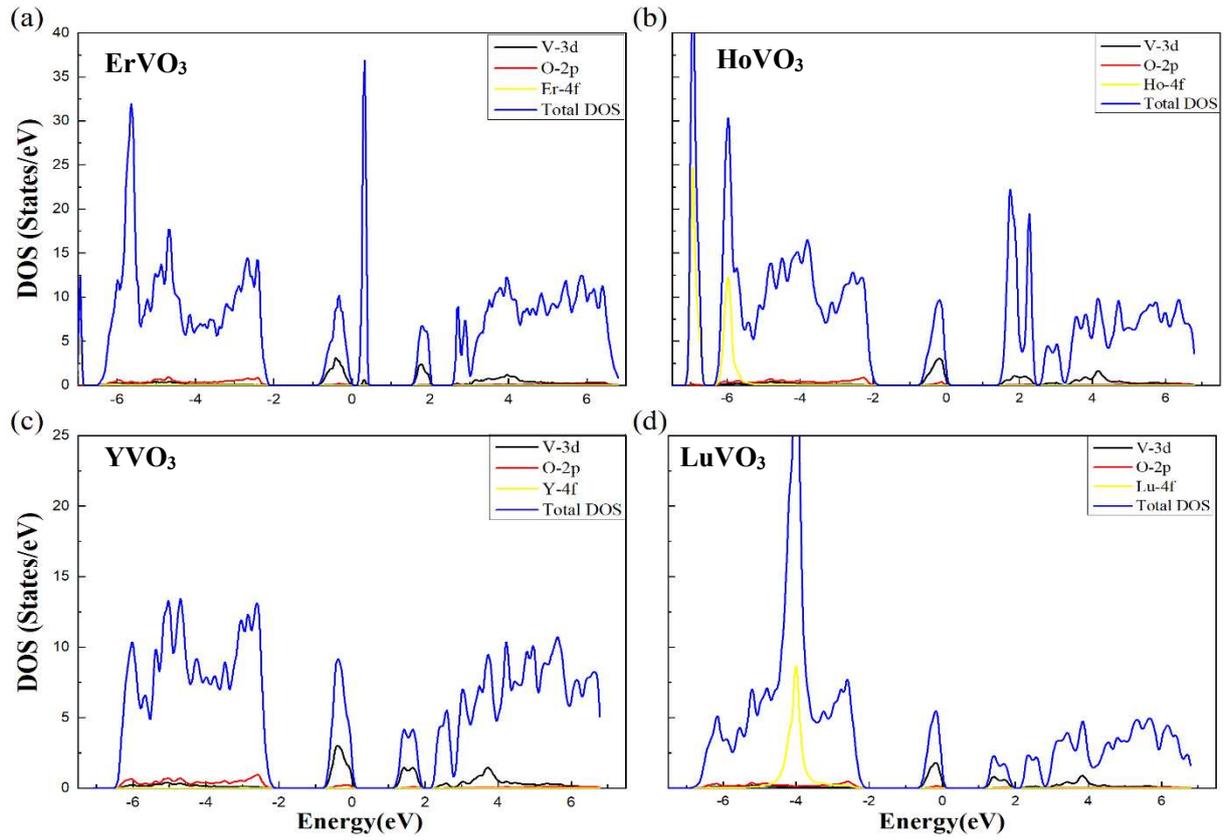

Figure 4: Total and partial density of states obtained within (GGA + U + spin orbit coupling) of. (a) ErVO$_3$. (b) HoVO$_3$ (c) YVO$_3$. (d) LuVO$_3$.

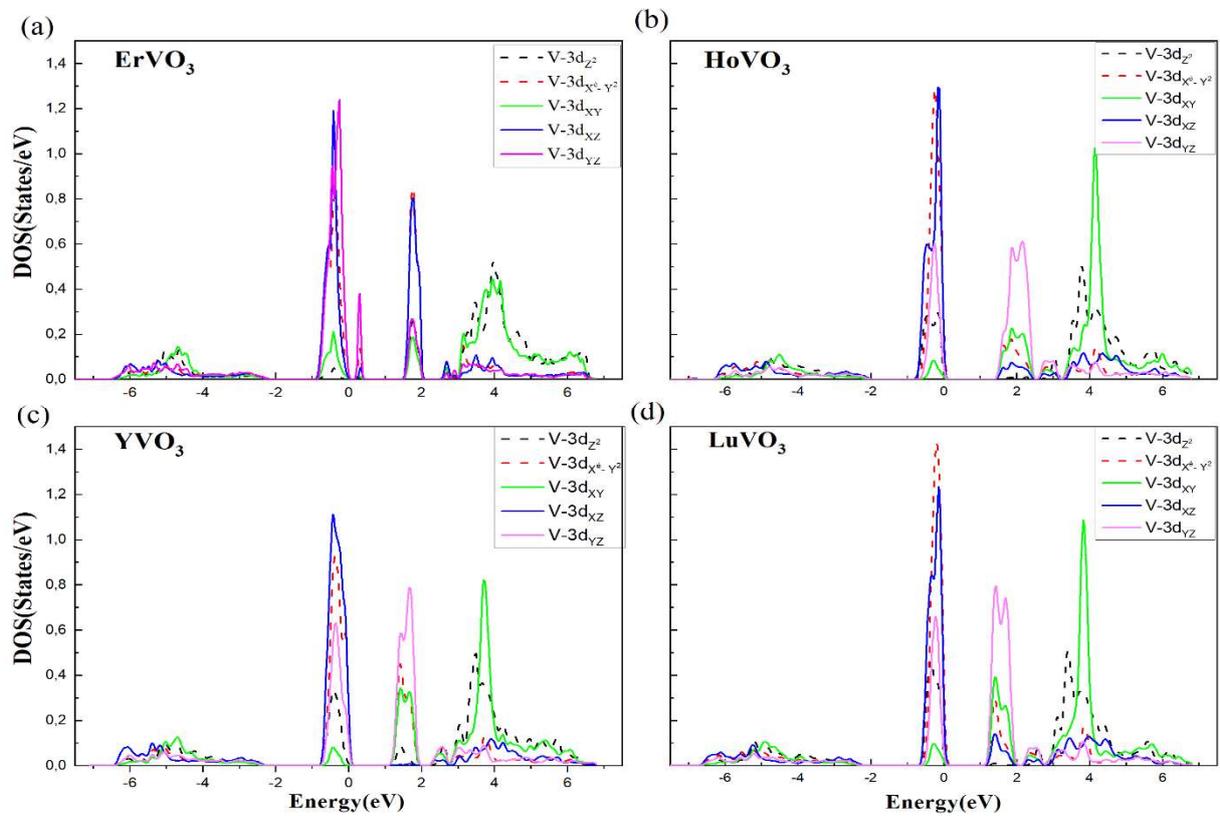

Figure 5: The partial density of vanadium d orbitals. (a) for ErVO$_3$, (b) for HoVO$_3$, (c) for YVO$_3$ and, (d) for LuVO$_3$.

The total magnetic moment of the elementary cell in the examined oxides, which contain four rare earth atoms, 4 vanadium atoms, 4 oxygen atoms of the first type, and 8 oxygen atoms of the second type, is almost zero for all compounds, because of their antiferromagnetism. The spin moments in the atomic muffin-tin spheres are summarized in table 3:

Table 3: Spin moments for Er, Ho, Y, Lu, V, and O in $ErVO_3$, $HoVO_3$, $YVO_3$, and $LuVO_3$.

| Compounds | $ErVO_3$ | | | | $HoVO_3$ | | | | $YVO_3$ | | | | $LuVO_3$ | | | |
|---|---|---|---|---|---|---|---|---|---|---|---|---|---|---|---|---|
| | Er | V | O1 | O2 | Ho | V | O1 | O2 | Y | V | O1 | O2 | Lu | V | O1 | O2 |
| Spin moments ($\mu_B$) | 2.83 | 1.42 | 0.05 | 0.05 | 3.83 | 1.57 | 0.002 | 0.002 | 0.002 | 1.6 | 0.001 | 0.001 | 0.01 | 1.61 | 0.01 | 0.01 |

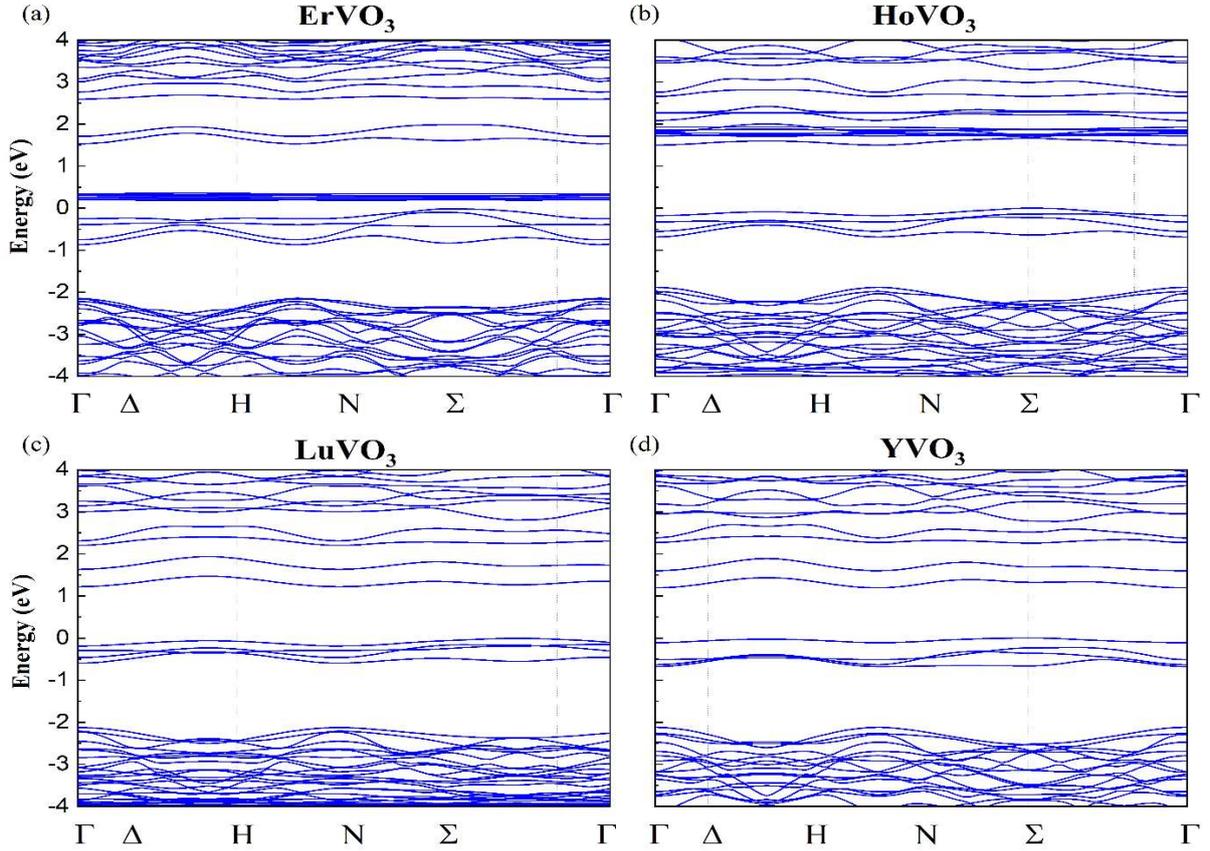

Figure 6: The band structure within (GGA + U + spin orbit coupling) for (a) $ErVO_3$, (b) $HoVO_3$, (c) $LuVO_3$ and, (d) $YVO_3$.

According to the spin moments, and the density of state calculations, holmium 4f electrons are responsible for most of the magnetic moment generation in $HoVO_3$ at holmium atoms, while the Ho d-electrons play a minor role, and electrons with s- and p- symmetry have no contribution. In contrast, the magnetic moment of vanadium atoms arises mostly from the strong polarization of $3d_{yz}$ and $3d_{z^2}$ states, and less from the polarization of $3d_{x^2-y^2}$ and $3d_{xz}$ orbitals. On the other hand, vanadium $3d_{xy}$ electrons are mostly inactive when it comes to the creation of the magnetic moment. The spin moment distribution among individual atoms in $ErVO_3$ is qualitatively similar to that in $HoVO_3$. Also, the obtained results indicate that the charge states of vanadium and oxygen atoms in holmium and erbium vanadates are almost similar. However, in vanadium atoms, there are certain differences in the occupancy numbers of specific d-harmonics.

### 3.3. Magnetic properties

In order to better understand the magnetic interactions in the studied $RVO_3$ compounds, the spins and orbital moments of rare earth elements R and V for $HoVO_3$, and $ErVO_3$ are estimated. The relative direction of magnetic moments between them and with respect to the applied external field is determined by the sign of XMCD signal and its integral [40].

In the Self-Consistent Field (SCF) computation, distinguishing between contributions from spins and orbitals is more difficult because the total magnetic moment is similar to the difference between up and down spin densities of state. Consequently, XMCD can provide additional information about the value of spin-orbit coupling and estimate the spin and orbital moments separately [40].

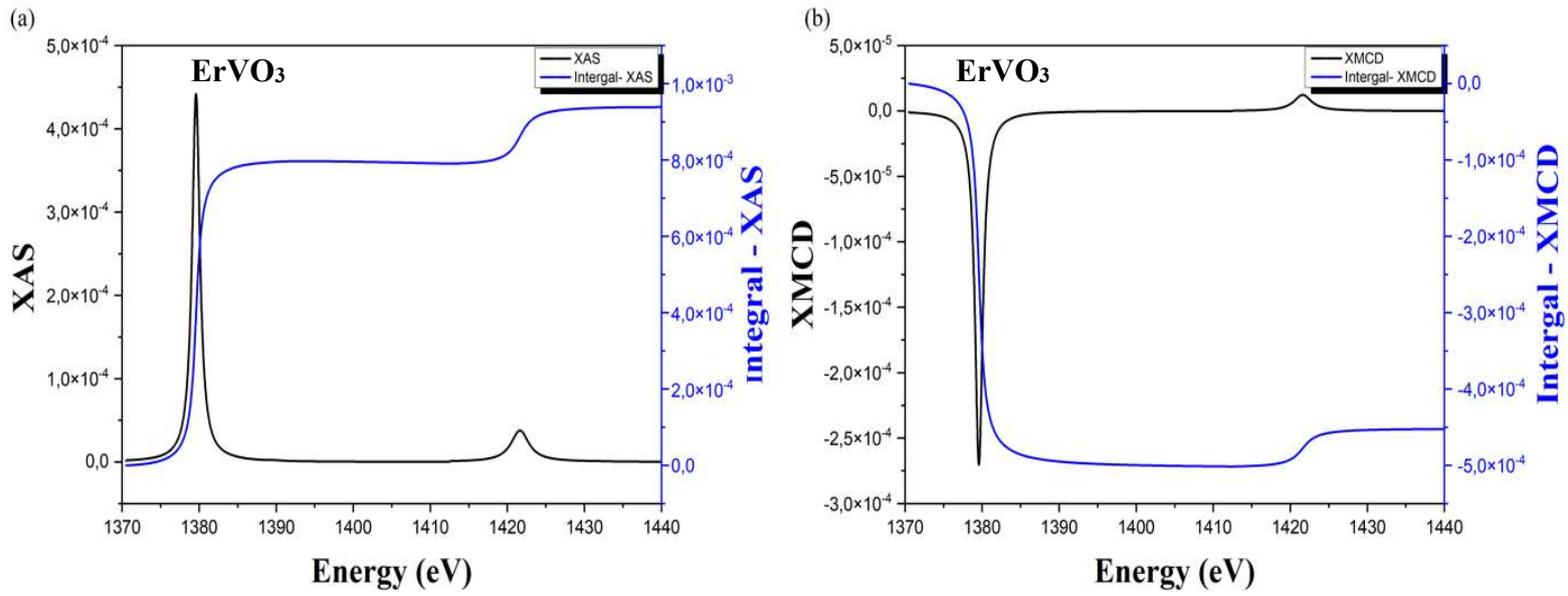

Figure 7: (a) XAS, and (b) XMCD spectra of Erbium atom in ErVO$_3$ representing M4,5 (black line) and the integral of the spectra (blue line) within GGA+U approximation.

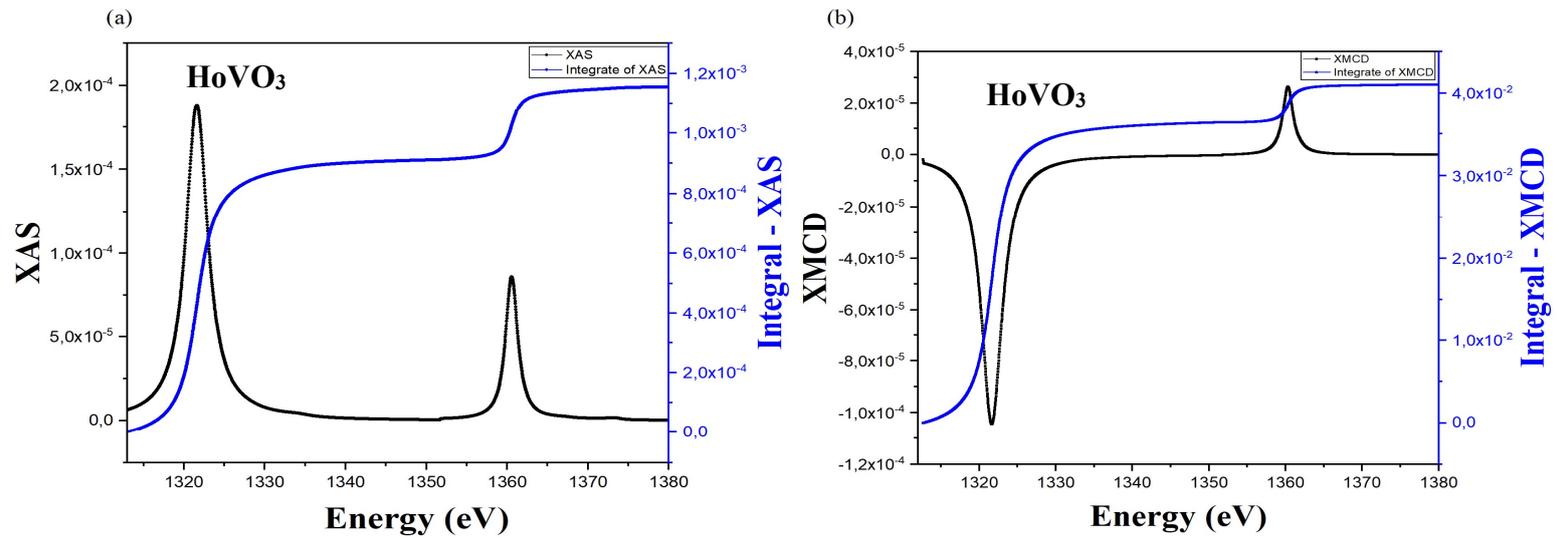

Figure 8: (a) XAS, and (b) XMCD spectra of Holmium atom in HoVO$_3$ representing M4,5 (black line) and the integral of the spectra (blue line) within GGA+U approximation.

The M4,5 edge is the absorption edge that offers the most direct information on rare earth 3d-4f transitions. Except for systems where the rare earth shows valence instability or mixed valence, the theoretical study of that edge is often done by using a model in which interactions with the environment, such as the crystal field, are considered to be of mild perturbation [49]. Due to localized rare earth element 4f electrons, this technique is valid, as there is no overlap between the 4f levels of two neighboring atoms [49].

M$_4$ and M$_5$ are separated in energy by the significant spin-orbit effect in Er, and Ho (3d-4f) absorption spectra seen in Figures 7 and 8. When compared to the M$_4$ edge, the intensity at the

$M_5$ edge is quite high, indicating a significant contribution from the orbital moment. For this purpose, both the spin and orbital moments must be calculated. In this way, the predicted value of a free $Er^{3+}$ ($4f^{11}$) magnetic moment regarding Hund's rule ground state (J = 15/2, S = 3/2, and L = 6) is 9.58 $\mu_B$ with J and L are the orbital and the total angular moments, respectively. For $Ho^{3+}$ ($4f^{10}$) the obtained magnetic moment (J = 8, S = 2, L = 6) is 10.6 $\mu_B$ assuming the isolated atoms. Whereas it is necessary to consider the atomic environment. For this purpose, the XMCD, and XAS technics are used. As shown in Figures (7 and 8), the negative (positive) of XMCD, and XAS signals indicates that the Erbium and Holmium magnetic sublattices are aligned parallel to the applied magnetic field and may be relate to the quadrupole dual transition: 3d to 4f and 4f to 5s. Figures 7, and 8 show the integral of XAS and XMCD, which allows us to compute the spin and orbital magnetic moments.

Table 4: Spin, orbital, and total magnetic moments for Er, Ho, and V atoms in (Er, Ho)VO$_3$.

|  | XMCD | | Experimental | | XMCD | | Experimental | |
| --- | --- | --- | --- | --- | --- | --- | --- | --- |
|  | Er | V | Er | V | Ho | V | Ho | V |
| Spin moment ($\mu_B$) | 2.82 | 1.42 | - | - | 3.83 | 1.57 | - | - |
| Orbital moment ($\mu_B$) | 5.19 | 0.032 | - | - | 2.94 | 0.019 | - | - |
| Total moment ($\mu_B$) | 8.01 | 1.452 | 8.2 [25] | 1.47 [25] | 6.77 | 1.58 | 6.9 [25] | 1.42 [25] |

We also attempted to obtain detailed information on the magnetic contribution of the V ions. The reported results in Table 4 reveal that only a low magnetic moment is produced within the antiferromagnetic cycloidal arrangement, indicating that the V lattice has a minor impact on total magnetization along the easy axis.

For magnetic materials containing heavy atoms like Ho or Er, the spin-orbit coupling (SOC) is crucial because it can generate magnetocrystalline anisotropy. To further understand the magnetic anisotropy of RVO$_3$, the DIPAN package in WIEN2k is used to determine the compound energy along each axis. The process requires calculating total energy for various axes in order to identify the lowest energy. For this purpose, the total energies of orthorhombic

HoVO$_3$, ErVO$_3$, YVO$_3$, and LuVO$_3$ are estimated using the GGA+U+SOC technique by orienting the magnetization following different directions. The obtained results are summarized in table 5.

Table 5: The total energies of RVO$_3$ (R = Ho, Er, Y, Lu) using GGA+U with different directions of magnetization by taking the SOC (spin orbit coupling) into account.

| Direction of magnetization | [100] | [110] | [010] | [011] | [101] | [001] | [111] |
|---|---|---|---|---|---|---|---|
| Total energy of HoVO$_3$ (J/$m^3$) | -0.3935559.10$^5$ | -0.3968426.10$^5$ | -0.3997389.10$^5$ | 0.3731598.10$^5$ | 0.4158956.10$^5$ | 0.7932948.10$^5$ | 0.1953113.10$^5$ |
| Total energy of ErVO$_3$ (J/$m^3$) | -0.2913005.10$^5$ | -0.2421789.10$^5$ | -0.1308839.10$^5$ | 0.9907150.10$^4$ | 0.1223876.10$^5$ | 0.1431074.10$^5$ | 0.1981675.10$^5$ |
| Total energy of YVO$_3$ (J/$m^3$) | 0.3470233.10$^4$ | 0.5302542.10$^4$ | 0.6937247.10$^4$ | -0.4296704.10$^4$ | -0.5873504.10$^4$ | -0.1040748.10$^5$ | -0.2439421.10$^4$ |
| Total energy of LuVO$_3$ (J/$m^3$) | 0.3723629.10$^4$ | 0.5403959.10$^4$ | 0.6887676.10$^4$ | -0.4415755.10$^4$ | -0.5944888.10$^4$ | -0.1061130.10$^5$ | -0.4415755.10$^4$ |

As reported, the lowest energy is clearly along the [001] direction for LuVO$_3$, and YVO$_3$. These results indicate that the easy magnetization axis is along the c-axis for these two compounds. These findings are in a good agreement with reported experimental data in Ref. [50]. The easy magnetization axis is located along the b-axis for HoVO$_3$. Based on the values of the total energy values, the easy magnetization direction might be along the ab plane. In ErVO$_3$, the easy magnetization axis is located along the a-axis. The same assumption can be considered in the case of ErVO$_3$. This demonstrates the non-collinearity of magnetic atom spins in (Er, Ho)VO$_3$ [51]. The observed changes concerning the easy axis between the HoVO$_3$ and ErVO$_3$ compounds is probably related to the difference in exchange interactions between the vanadium and rare earth lattices, as well as the orbital-lattice coupling (the existence of Jahn Teller effect) [48]. These results are in good agreement with the reported experimental data in Refs. [28], and [50]. These findings underline the fact that the nature of the rare earth elements (non-magnetic or magnetic) might markedly control the magnetization easy direction as well as the strength of the magnetic anisotropy in RVO3 orthovanadates. Such a magnetocrystalline anisotropy would lead to a giant rotating magneto-caloric effect for the vanadate oxides family [52,53] that can be obtained by spinning their single crystals between their easy and hard magnetization axes without changing the external magnetic field. For example, a large RMCE can be obtained by rotating YVO3 and LuVO3 crystals within the bc plane in constant magnetic fields. In addition, the rotating magnetocaloric effect in materials containing magnetic rare earth elements (Er, Ho)VO3 may be created with performance similar to the conventional magnetocaloric effect found in HoVO3 [28], by rotating these crystals within the ac plane or bc plane for ErVO3, and HoVO3, respectively.

### *3.4. Exchange coupling in RVO3 with (R=Er, Ho)*

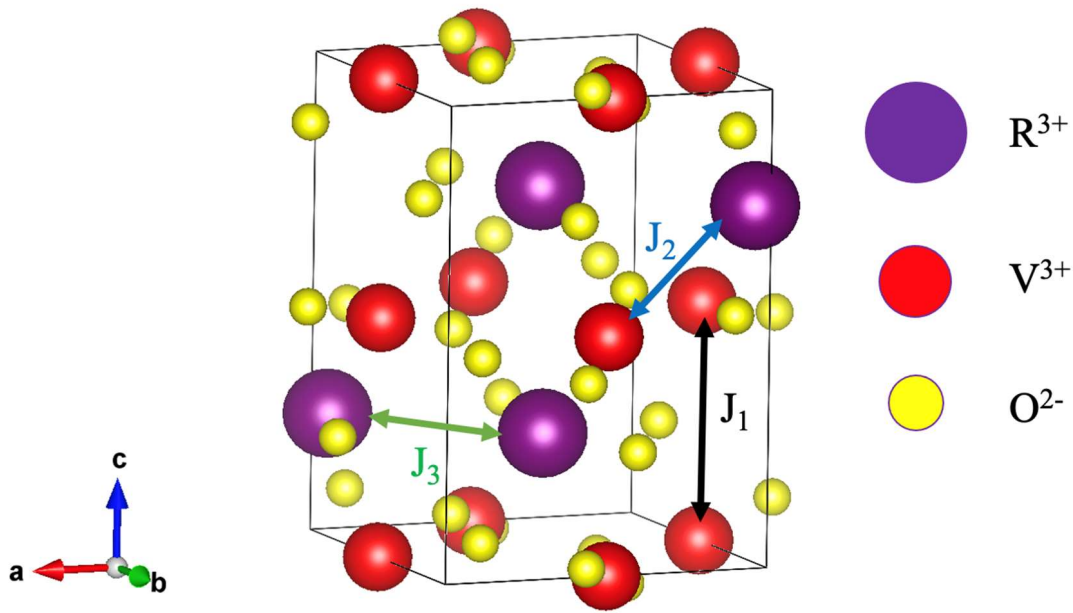

Figure 7: Exchange coupling interactions. J1 is the exchange coupling between(V-V), J2 is the exchange coupling between (V-R), and J3 is the exchange coupling between (R-R).

The exchange coupling interactions between atoms is another key parameter that enable us to understand the driving mechanisms behind magnetic properties of materials. By mapping the density fuctional theory (DFT) total energy to the Ising model [54], the magnetic exchange interactions can be calculated by using the Hamiltonian as follow:

$$H = -\sum_{ij} -J_{ij} S_i S_j \quad (3)$$

Where H is the Hamiltonian, $J_{ij}$ is the exchange interaction between $S_i^z$ and $S_j^z$, and $S_i, S_j$ are the spins at the site i and j, respectively. Here, three independent exchange interactions are considered where $J_1$, $J_2$ and $J_3$ are the exchange interactions between V atoms and R (R = Er, Ho)-V, and R atoms, respectively. The exchange coupling interactions are illustrated in Fig. 9. At site i, $S_i$ is the spin operator. The spin exchange parameter $J_{ij}$ is confined to the nearest (R–V) and next nearest (R-R and V-V) spin pairs, and the summation is over spin pairs. The spin magnetic moments for HoVO$_3$ in the spheres of Ho and V are 3.85 $\mu_B$ and 1.57 $\mu_B$, respectively. In ErVO$_3$, they are found to be 2.82 $\mu_B$ and 1.53 $\mu_B$ for Er, and V, respectively. The $Ho^{3+}$, $Er^{3+}$, and $V^{3+}$ cations should theoretically contribute for 4 $\mu_B$, 3 $\mu_B$, and 2 $\mu_B$, respectively. The Ho, Er, and V spins were taken s = 2, s = 3/2, and s = 1, respectively. To reduce the errors, we work with the total energy differences with respect to the ground state rather than the absolute total-energy values. The orthorhombic structure of (Er, Ho)VO$_3$ was used as a reference. Then, the magnetic order was switched between four different spin configurations (ferromagnetic, and three antiferromagnetic types) and calculated their total energy. Therefore, the four total energies of RVO$_3$ with (R = Er, Ho) are given by:

$$\begin{cases} E(C-AFM) = J_1 \sum_{<i,j>} S_i S_j - J_2 \sum_{<i,j>} S_i \sigma_j + J_3 \sum_{<i,j>} \sigma_i \sigma_j \\ E(G-AFM) = J_1 \sum_{<i,j>} S_i S_j + J_2 \sum_{<i,j>} S_i \sigma_j - J_3 \sum_{<i,j>} \sigma_i \sigma_j \\ E(A-AFM) = -J_1 \sum_{<i,j>} S_i S_j + J_2 \sum_{<i,j>} S_i \sigma_j + J_3 \sum_{<i,j>} \sigma_i \sigma_j \\ E(Ferro) = -J_1 \sum_{<i,j>} S_i S_j - J_2 \sum_{<i,j>} S_i \sigma_j - J_3 \sum_{<i,j>} \sigma_i \sigma_j \end{cases} \quad (4)$$

Table 6: The exchange coupling values of HoVO$_3$ and ErVO$_3$.

| Exchange coupling (meV) | $J_1$ | $J_2$ | $J_3$ |
|---|---|---|---|
| HoVO$_3$ | - 5.46 | - 14.83 | 7.49 |
| ErVO$_3$ | - 13.42 | - 3.82 | 2.41 |

The energies parameters can be calculated using the preceding equations, and then extract the exchange coupling parameters $J_{ij}$. Accordingly, in HoVO$_3$ the spin exchange parameters $J_{ij}$ for the closest V-V pair are -5.46 meV, -14.83 meV for Ho-V, and 7.49 meV for Ho-Ho. Therefore, we notice that the Ho-V spin coupling is much greater than the other two exchange couplings, indicating a strong interaction between the rare earth element and the transition metal element in this material. On the other hand, in ErVO$_3$, the spin exchange parameters $J_{ij}$ for the closest V-V pair are - 13.42 meV, - 3.82 meV for Er-V, and 2.41 meV for Er-Er indicating that the exchange coupling V-V is the strongest for ErVO$_3$. The ionic radius of the $R^{3+}$ ions might influence these interactions [25]. However, the magnetic structure of HoVO$_3$ is much more complicated due to the interplay between the Jahn-Teller effect and the competing V-V, Ho-V, and Ho-Ho exchange interactions [25]. Otherwise, for ErVO$_3$ only the V-V interaction dominate.

According to Miyassa et al. [55], the magnetic field H along the a and b axes produces the metamagnetic transition of the Dy Ising moments. In fact, the important Dy(4f) - V(3d) spin exchange interaction causes the rearrangement of V$^{3+}$ spins from G-type to C-type, and therefore the flipping of the pattern of occupied $d_{yz} = d_{zx}$ orbitals from C-type to G-type due to the alignment of Dy 4f moments under magnetic fields. Balli et al. [28] have experimentally observed the same phenomenon in HoVO$_3$ crystals. The above underlined important exchange interaction Ho(4f) - V(3d) causes the metamagnetic transition of the Ho Ising moments leading to a giant conventional magnetocaloric effect of about 17.2 J/kg K for 7 T. Thus, there is a possibility to generate a similar large magnetocaloric effect in DyVO$_3$ to the one found in HoVO$_3$ or higher, since both compounds exhibit the same electronic and magnetic behaviors. However, a lower MCE is expected in ErVO$_3$ compound due the weakness of Er-V exchange coupling when compared to HoVO$_3$. In addition, DyVO$_3$, HoVO$_3$, and ErVO$_3$ orthovanadates have the potential to generate a rotating magnetocaloric effect, because of the strong magnetocrystalline anisotropy in these materials.

## 4. CONCLUSION

To sum up, the structural, electronic, and magnetic properties of RVO$_3$ with (R = Er, Ho, Y, Lu) are theoretically examined in the frame of ab initio calculations utilizing the FPLAW method using GGA and GGA+U and by considering the spin-orbit coupling. The obtained results are consistent with the experimental data. The magnetic phase stability of RVO$_3$ (R=Er, Ho, Y, Lu) compounds has been explored within various magnetic configurations to figure out the most stable structure. Further, by using the density of state calculations, these compounds were found to exhibit a semiconducting behavior with band gaps equal to 0.58 eV, 1.57 eV, 1.12 eV, 1.2 eV, for ErVO$_3$, HoVO$_3$, YVO$_3$, and LuVO$_3$, respectively. The reported magnetic moments and magneto crystalline anisotropies are in good accord with earlier experimental observations. Furthermore, the exchange coupling between Ho-V is found to be greater than other exchange interactions in HoVO$_3$. In contrast, the most important exchange coupling in ErVO$_3$ occurs between vanadium atoms V-V. This would be explained by the difference in the ionic radius of the rare earth element. These obtained key parameters by DFT calculations are important for studying the magnetocaloric properties of $RVO_3$ (R = Er, Ho, Y, Lu). Therefore, the modeling of several critical magneto-thermal properties of $RVO_3$ materials, such as specific heat, entropy, and adiabatic temperature fluctuations, remains the primary issue. Our team is currently investigating this problematic, and any relevant findings will be presented in a future paper.

## Acknowledgement

The authors acknowledge funding by the International University of Rabat.
We would like also to thank Prof. H. Essadiqi for partly making possible the calculation by providing computational resources.
S. Ait jmal would like to thank H. ZAARI for helpful discussions and computations.

## Compliance with Ethical Standards:

Conflict of Interest: The authors declare that they have no conflict of interest.